\documentclass[runningheads]{llncs}
\usepackage{graphicx}
\usepackage{amsmath}
\usepackage{amsfonts}
\usepackage{amssymb}
\usepackage{url}
\urldef{\mailsa}\path|| \urldef{\mailsb}\path||

\begin{document}
\bibliographystyle{plain}
\title{Human Heuristics for Autonomous Agents}

\titlerunning{}

\vskip 1cm

\author{Franco Bagnoli\,$^{\rm 1}$\thanks{to whom correspondence
should be addressed},  Andrea Guazzini\,$^{\rm 1}$ and Pietro
Li\`{o}\,$^{\rm 2}$\\
{\small
$^{\rm 1}$Department of Energy, University of Florence, Via S. Marta
3, 50139 Firenze, Italy. Also CSDC and INFN, sez. Firenze}\\
{\small $^{\rm 2}$Computer Laboratory, University of Cambridge, 15
 J.J. Thompson Avenue, Cambridge, CB30FD, UK}\\
}
\authorrunning{Heuristics}
\institute{
\mailsa\\
\mailsb\\}

\maketitle

\begin{abstract}
We investigate the problem of autonomous agents processing pieces of
information that may be corrupted (tainted). Agents have the option of contacting a
central database for a reliable check of the status of the message,
but this procedure is costly and therefore should be used with
parsimony. Agents have to evaluate the risk of being infected, and
decide if and when communicating partners are affordable. Trustability
is implemented as a personal (one-to-one) record of past contacts
among agents, and as a mean-field monitoring of the level of message
corruption. Moreover, this information is slowly forgotten in time,
so that at the end everybody is checked against the database. We
explore the behavior of a homogeneous system in the case of a fixed
pool of spreaders of corrupted messages, and in the case of
spontaneous appearance of corrupted messages.
\end{abstract}

\section{Introduction}

One of the most promising area in computer science is the design of
algorithms and computer architectures closely based on our reasoning
process and on how the brain works. Human neural circuits receive,
encode and analyze the ``available information'' from the environment
in a fast, reliable and economical way. The evolution of human
cognition could be viewed as the result of a continuous improvement
of neural structures which drive the decision making processes from
the inputs to the final behaviors, cognitions and
emotions. Heuristics are simple, efficient rules, hard-coded by
evolutionary processes or learned, which have been proposed to
explain how people make decisions, come to judgments, and solve
problems, typically when facing complex problems or incomplete
information. It is common experience that that much of human
reasoning and decision making can be modeled by fast and frugal
heuristics that make inferences with limited time and knowledge. For
example, Darwin's deliberation over whether to marry provides an
interesting example of such heuristic
process~\cite{Healey,Litchfield1915}.

Let us quickly review some widely accepted hypothesis about
heuristics. In the early 1970s,
Daniel Kahneman and Amos Tversky (K\&T) produced a series of
important papers about decisions under
uncertainty~\cite{Tversky73,Tversky74,Tversky79,Tversky81,Tversky82}.
Their basic claim was that in assessing
probabilities, \emph{``people rely on a limited number of heuristic
principles which reduce the complex tasks of assessing probabilities
and predicting values to simpler judgmental operations''}.
Although K\&T claimed that, as a general rule,
heuristics are quite valuable, in some
cases, their use leads \emph{``to severe and systematic errors''}. One
of the
most striking features of their argument was that the errors follow
certain statistics and, therefore, they could be described and even
predicted. The
resulting arguments have proved highly influential in many fields,
including computer science (and particularly in human-machine
interaction area) where the influence has stemmed from the effort to
connect algorithmic accuracy to speed of elaboration and, equally
important, to the algorithmic
understanding of the human logic~\cite{Tversky82}. If human
beings use identifiable heuristics, and if they are prone to
systematic errors, we might be able to design computer architectures
and algorithms to improve human-computer interaction (and also to
study human behavior).

K\&T described three general-purpose heuristics:
\textbf{representativeness},
\textbf{availability} and \textbf{anchoring}. People use the
\emph{availability} heuristic when
they answer a question of probability by relying upon knowledge that
is readily available rather than examine other alternatives or
procedures. There are situations in which people assess the
frequency of a class or the probability of an event by the ease with
which instances or occurrences can be brought to mind. For example,
one may assess the risk of heart attack among middle-aged people by
recalling such occurrences among one's acquaintances. Availability
is a useful clue for assessing frequency or probability, because
instances of large classes are usually reached better and faster
than instances of less frequent classes. However, availability is
affected by factors other than frequency and probability. This is a
point about how familiarity can affect the availability of
instances. For people without statistical knowledge, it is far from
irrational to use the availability heuristic; the problem is that
this heuristic can lead to serious errors of fact, in the form of
excessive fear of small risks and neglect of large ones.

The \emph{representativeness} heuristic is involved when people make
an assessment of the degree of correspondence between a sample and a
population, an instance and a category, an act and an actor or, more
generally, between an outcome and a model. This heuristic can be
thought of as the reflexive tendency to assess the similarity of
characteristics on relatively salient and even superficial features,
and then to use these assessments of similarity as a basis of
judgment. Representativeness is composed by categorization and
generalization: in order to forecast the behavior of an (unknown)
subject, we first identify the group to which it belongs
(categorization) and them we associate the ``typical'' behavior of the
group to the item. Suppose, for example, that the question is
whether some person, Paul, is a computer scientists or a clerk
employed in the public administration. If Paul is described as shy
and withdrawn, and as having a passion for detail, most people will
think that he is likely to be a computer scientist and ignore the
``base-rate'', that is, the fact that there far more
clerk employed in public admin than computer scientists. It should
be readily apparent that the representativeness heuristic will
produce problems whenever people are ignoring base-rates, as they
are prone to do.

K\&T also suggested that estimates are
often made from an initial value, or \emph{anchoring}, which is then
adjusted to produce a final answer. The initial value seems to have
undue influence. In one study, K\&T asked subjects to say whether the
number that emerged from the wheel was higher or lower than the
relevant percentage. It turned out that the starting point, though
clearly random, greatly affected people's answers. If the starting
point was 65, the median estimate was 45\%; if the starting point
was 10, the median estimate was 25\%.

Several of recent contributions on heuristic have put the attention
on the ``dual-process''  to human
thinking~\cite{McNeil82,Taylor1982,Slovic1982,Gigerenzer1999,%
Christensen2000}.
According to these hypothesis, people have two systems for making
decisions. One of them is rapid, intuitive, but sometimes
error-prone; the other is slower, reflective, and more statistical.
One of the pervasive themes in this collection is that heuristics
and biases can be connected with the intuitive system and that the
slower, more reflective system might be able to make corrections.
The dual-process idea has some links with the experimental evidences
of the presence of areas for emotions in the brain, for instance of
fear-type. These ``emotional'' areas may be triggered before than the
cognitive areas become involved.

We shall try to consider some of
these concepts to model autonomous agents that have the  task of
processing messages from
sources that are not always trustable. The agent is a direct
abstraction of an human being, easily
understandable by psychologists and biologist with the advantage of
following a stochastic dynamics that can be combined with other
approaches like
ODE~\cite{Weiss2000,Wooldridge2002,Dinverno2003,Sun,Merelli2007}.
Here we make the analogy between the diffusion of hoaxes, gossips,
etc., and that of computer viruses or worms.

The incoming information
may be corrupted for many reasons:
some agents may be infected by malware and particularly viruses,
some of them may be programmed to provide false information or they
may just be malfunctioning. Let us suppose that  the
processing of a corrupted information will infect the elaborated
message, so that the corruption ``percolates and propagates'' into the
connection network, unless stopped. We assume that an agent may
contact a central database for inquiring about the reliability of a
message, but this checkout is costly, at least in terms of the time
required for processing the information. Therefore, an agent is
confronted with two opportunities: either trust the sender, accept
the message and the risk or passing false information and process it
in a short time, or contact the central database, be sure of the
correctness of the message but also waste more time (or other
resources such as bandwidth) in elaborating it. This is analogous
to the passport check when crossing a boundary: customers may
either trust the identity card and let people pass quickly, or
check them against a database, slowing down the queue.

This paper, which is motivated by the fact that human heuristics may
be used to improve the efficiency of artificial systems of
autonomous decision-makers agents, is structured as follows. In
Section~\ref{model}, we introduce a model where the above mentioned
heuristics are implemented.  Section \ref{relaxation} focuses
on equilibrium and asymptotic conditions in the absence of infection.
In Section~\ref{scenarios}, we describe
the different scenarios which are considered (no infection, quenched
infection and annealed infection); numerical results for
different value of control parameters under infection are reported
in Section~\ref{results}. A discussion about the psychological
implications of the model and conclusions are drawn in
Section~\ref{conclusions}.

\section{Model}\label{model}

Let us consider a scenario with $N$ agents, identified by the index
$i=1, \dots, N$. Each agent interacts with other $K$ randomly chosen
agents. The connections indicate messages transferred.  In
principle, one can have input connections with himself (meaning
further processing of a given piece of information) and multiple
connections with a given partner (more information transferred). An
agent receives information from its connecting inputs, elaborates it
and send the result to its output links. Let us assume for
simplicity that this occurs in a synchronous way and at discrete
time steps $t$. The information however can be tainted (corrupted),
either maliciously (virus, sabotage, attack) or because it is based
on incorrect data.

If an information is tainted, and it is accepted for
processing, it contaminates the output. All agents have the
possibility of
checking the correctness of the incoming messages against a central
database, but this operation is costly (say, in terms of time), and
therefore heuristics are used to balance between cost  and the risk
of being infected.

An agent $i$ has a dynamical memory for the reliability of its
partners $j$, $-1\le \alpha_{ij}\le 1$; this memory is used to decide if a message
is acceptable or not. The greater  $\alpha_{ij} > 0$,  the more the
partner is considered reliable, the reverse for $\alpha_{ij} < 0$.
However, the trusting on an individual is not an absolute value, it has
to be compared with the perception of the level of the
infection. Let us denote by $0\le A_i \le 1$ the perception of
the risk \emph{i.e.}, the perceived probability of message
contamination,
of individual $i$. A simple yet meaningful way of combining risk
perception with uncertainty is to assume that each individual $i$
decides according with its previous knowledge ($\alpha_{ij}$) if
$|\alpha_{ij}| > A_i$ and checks against the database (\emph{i.e.},
get to
know the truth) otherwise. If $A_i$ is large, the agent $i$ will be
suspicious and check many messages against the database, the reverse
for small values of $A_i$.

After checking the database, one knows the truth about his/her
partner. This information can be used to increase or decrease
$\alpha_{ij}$ and also to compute $A_i$. In particular, if the check
is positive (negative), $\alpha_{ij}$ increases (decreases) of a
given amount $v_\alpha$. Finally $A_i$ in increased by a quantity
$v_A n_i/c_i$, where $c_i$ is the cost (total number of checks for a
given time step) and $n_i$ the number of infected discovered. The
idea is that $A_i$ represents the perceived ``average'' level of
infection, corresponding to the ``risk perception'' of being
infected. We shall limit here to fixed and homogeneous responses, in
an more realistic case, different classes of agents or individuals
will react differently, according to their ``programming'' and their
past experiences, to a given perception of the infection level.

Some of these quantities change smoothly in time. There is an oblivion
mechanism on $\alpha_{ij}$ and $A_i$, implemented with the
parameters $r_\alpha$ and $r_A$, respectively, such that the
information stored $\tau$ time steps before the present time has
weight $(1-r)^\tau$. New information is stored with weight $r$. This
mechanism emulates a finite memory of the agent, without the need of
managing a list. 

The observable quantities are the total number of infected
individuals, $I$, the cost of querying the database, $C$ and the
number of errors $E$, which are given by the number of tainted
accepted messages and not-tainted refused messages.

In this model, we are only interested in the correctness of the
message, not in its content. Actually, a real message should be
considered as a set of 'atomic' parts, each of which can be
analyzed, eventually with their relations, in order to judge the
reliability of the message itself. For instance, the spam detection
mechanism is often based on a score assigned to patterns
(\emph{e.g.}, MONEY, SEX, LOTTERY) appearing in the message.
Therefore, a more accurate model should represent messages as
vectors or lists of items. We deal here with a simple scalar
approximation.

We try to include the human heuristics in this simple model by means
of $A$ (representativeness) and $\alpha_{ij}$ (availability).
The oblivion mechanism can moreover be considered the parameter
corresponding to the ``anchoring'' experiences.
In our present model, there is only one variable connected to 
affordability (from completely trustable to
completely not trustable), and the categorization procedure consists
essentially in trying to assess the placement of an individual on
this axis.  The trustability of an individual ($\alpha_{ij}$)
depends on the past interactions. Since $A$ represents the average
level of infectivity, the trustability of an individual is
evaluated against it, in order to save the cost (or the time) of the
check against the central database.

\section{Relaxation to equilibrium and asymptotic state without infection}
\label{relaxation}

\begin{figure}
 \includegraphics[width=0.32\columnwidth]{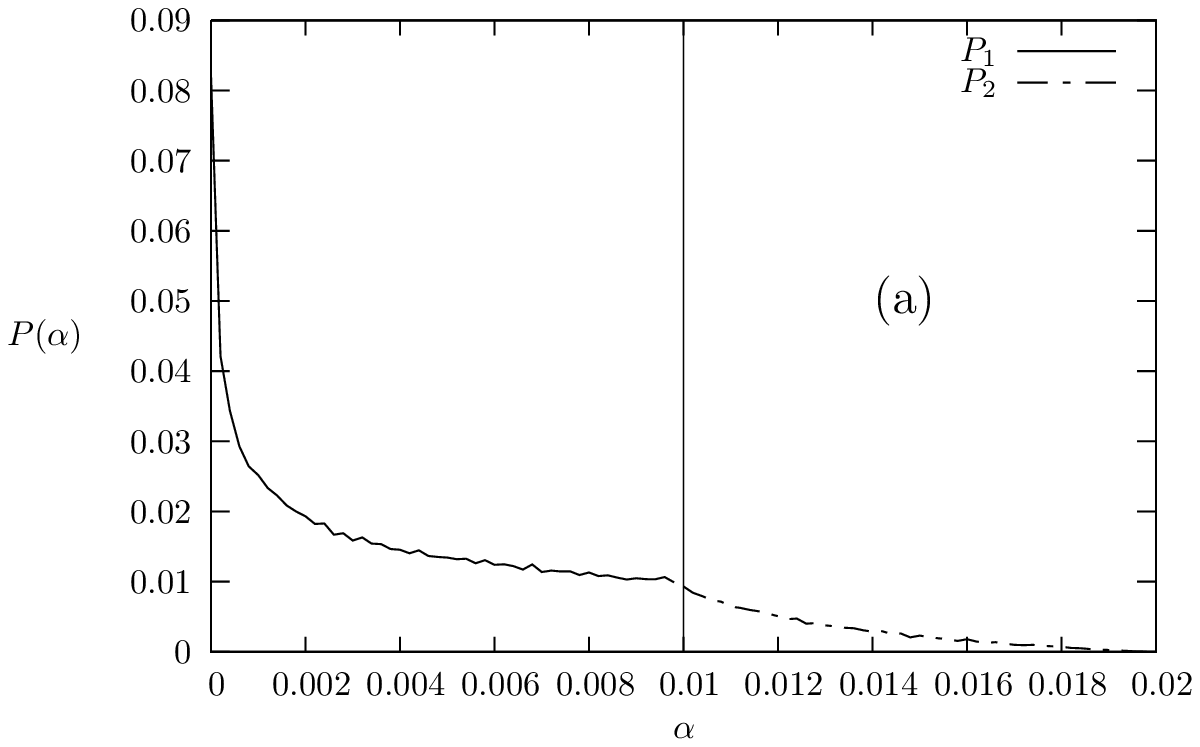}
 \includegraphics[width=0.32\columnwidth]{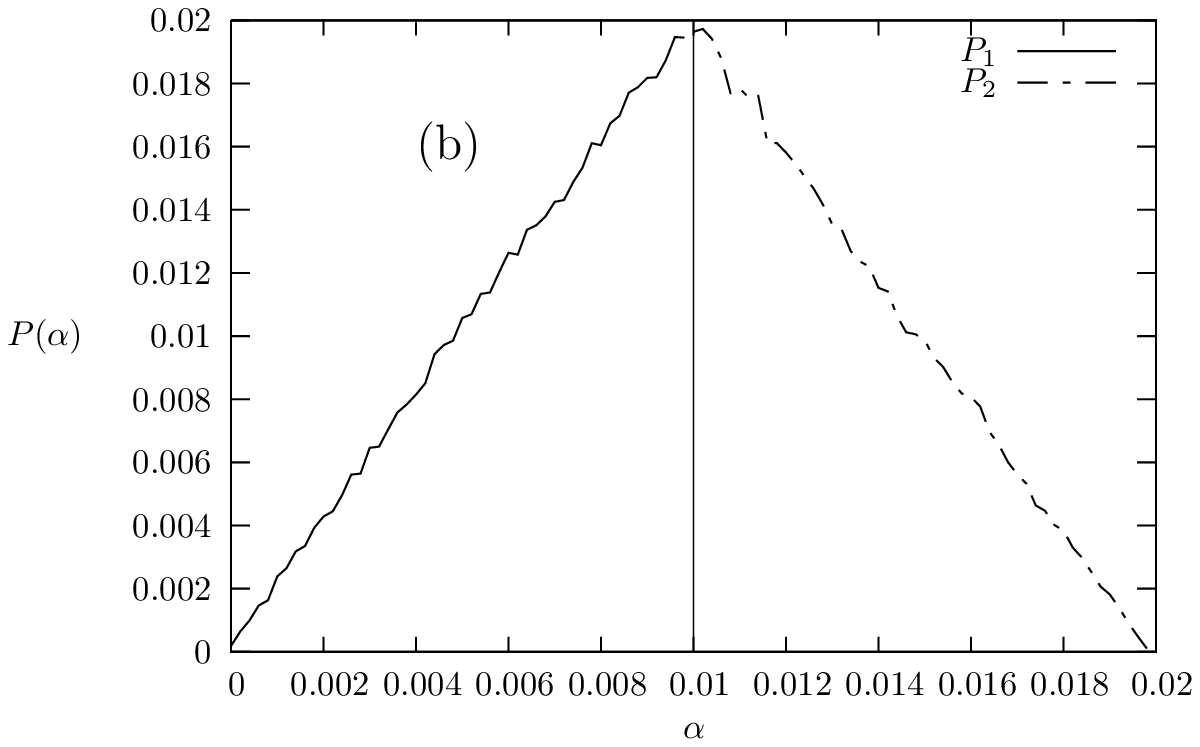}
 \includegraphics[width=0.32\columnwidth]{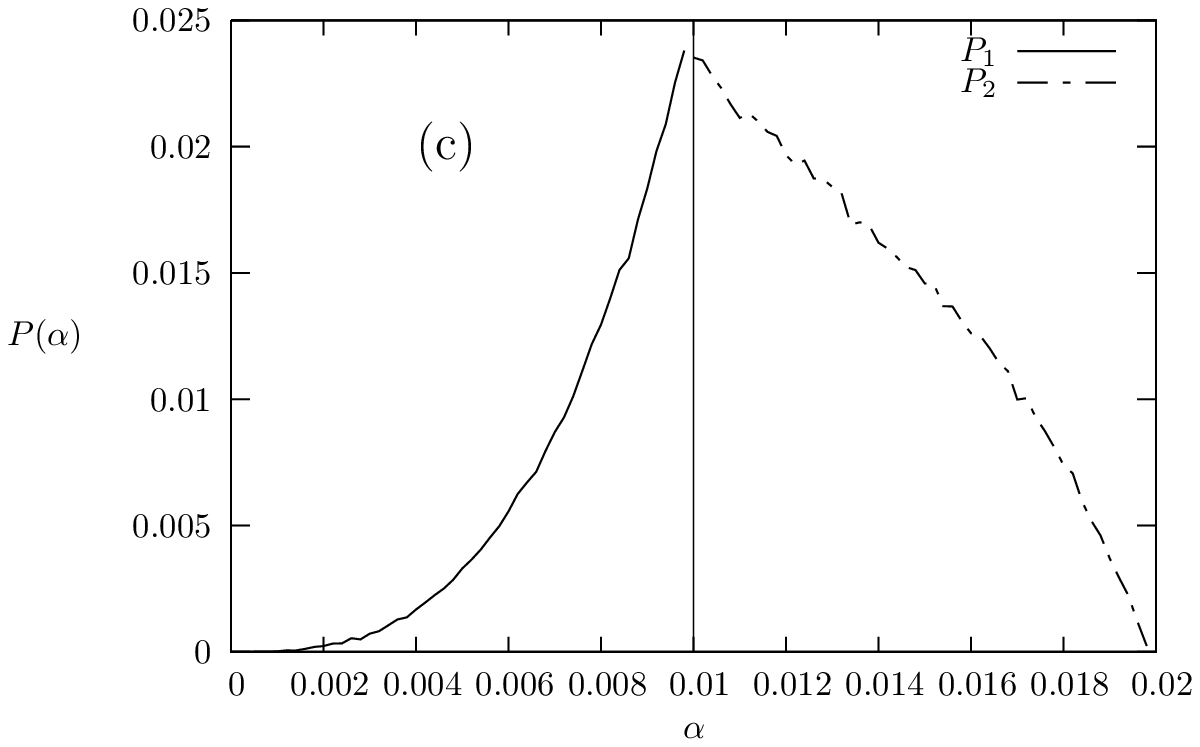}
\caption{\label{alpha}The asymptotic distribution $P(\alpha)$ for
$a<2r$ ($a=0.006$ and $r=0.01$) (a); $a=2r$ ($a=0.01$ and $r=0.005$) (b); $a>2r$ ($a=0.02$ and
$r=0.005$) (c).}
\end{figure}

In order to put into evidence the emerging features of our model, let
us
first study the case without infection. Without ``stimulation'', the
threshold  $A_i$ is
fixed, and takes the value $v_A$ for all individuals. The only
dynamical variables are the $\alpha_{ij}$.

Starting from a peaked (single-valued) distribution of $\alpha_{ij}$,
the model exhibits oscillatory patterns and long transients towards
an equilibrium distribution (Fig.~\ref{alpha}). We found that
by increasing the connectivity $K$, the peaks become thinner and
higher,  following a linear relationship. The affinities $\alpha_{ij}$
in the asymptotic state
have a non trivial distribution, ranging from $0$ to $2v$. Let us
call $P(\alpha)$ the probability distribution of $\alpha$. From
numerical simulation (see Fig.~\ref{alpha}), one can see that
$P(\alpha)$ can be divided into two branches, $P_1(\alpha)$ for
$0\le\alpha\le v$ and $P_2(\alpha)$ for $v\le\alpha\le 2v$. The
evolution of $P(\alpha)$ is given by the combination of two phases:
control against the database, that in the mean field approach occurs
with probability $a=K/N$ for all $\alpha \le v$ (and therefore for
$P_1$), and the oblivion mechanism, that multiplies all $\alpha$ by
$(1-r_\alpha)$. Combining the two effects, one finds for the
asymptotic state
\begin{align}
 P_1(\alpha) &= \frac{1-a}{1-r_\alpha}
P_1\left(\frac{\alpha}{1-r_\alpha}\right),\label{P1}\\
 P_2(\alpha) &=
\frac{a}{1-r_\alpha}P_1\left(\frac{\alpha}{1-r_\alpha}-v\right)+
\frac{1}{1-r_\alpha}P_2\left(\frac{\alpha}{1-r_\alpha}\right).
\end{align}

From Eq.~\eqref{P1}, one gets easily that $P_1(\alpha) \propto
\alpha^x$, with
\[
  x = \frac{\ln(1-a)}{\ln(1-r_\alpha)}-1 \simeq \frac{a}{r_\alpha}-1.
\]

In particular, the value $x=1$ (Fig.~\ref{alpha}-b) corresponds to
$a=2r_\alpha$. We were not able to express the asymptotic
distribution $P_2(\alpha)$ in terms of known functions.

\begin{figure}
 \includegraphics[width=0.5\columnwidth]{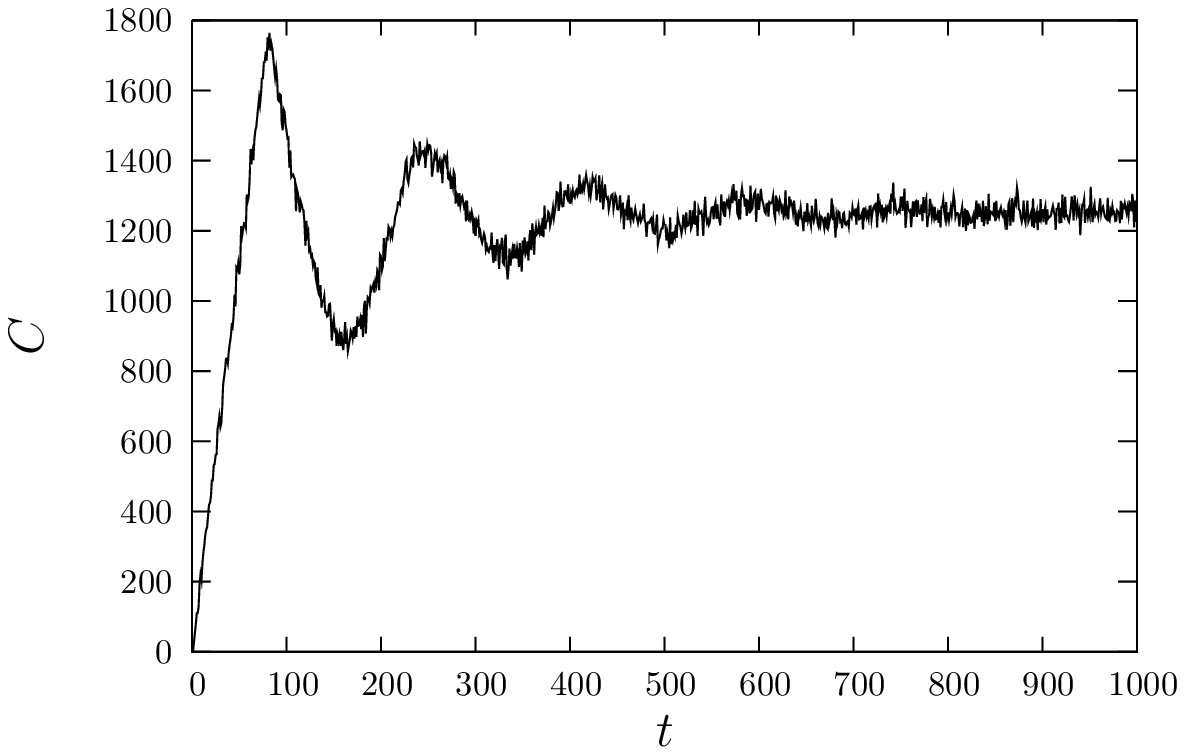}
 \includegraphics[width=0.5\columnwidth]{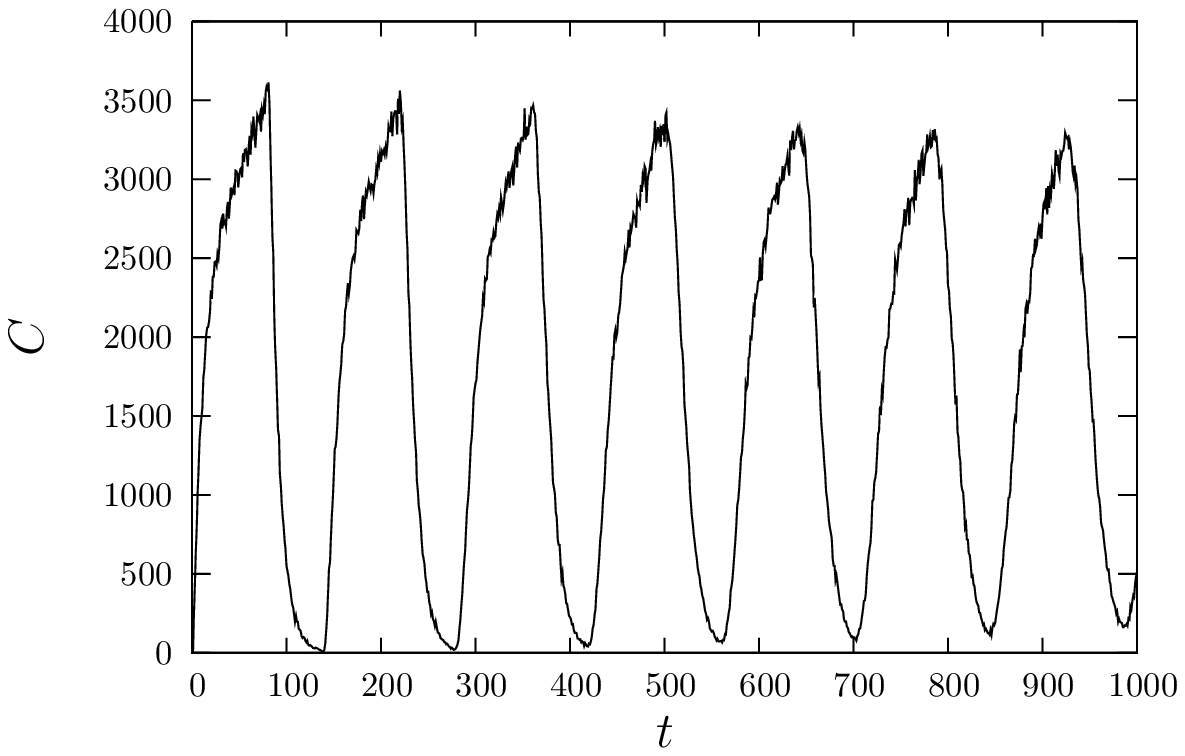}
\caption{\label{C-t} Relaxation to equilibrium for the cost $C$ for
$N=500$, $r_\alpha=0.005$ and $K=5$ ($a=0.01$) (left); $K=50$
($a=0.1$) (right).}
\end{figure}

The process of relaxation to the equilibrium is in general given by
oscillations, whose period is related to $r_\alpha$. A rough
estimation can be obtained by considering that a pulse of agents
with the same value of $\alpha=2v$ will experience the oblivion at an
exponential rate $(1-r_\alpha)^T$, until $\alpha=v$, after which a
fraction $a$ of the pulse is re-injected again to the value
$\alpha=2v$. The condition for the pseudo-periodicity (for the fraction $a$ of
agents) is
\[
  2v (1-r_\alpha)^T = v,
\]
from which the period $T$ can be estimated
\[
  T \simeq -\frac{\ln(2)}{\ln(1-r_\alpha)} \simeq
\frac{\ln(2)}{r_\alpha}
\]
in the limit of small $r_\alpha$.

Since the re-injected fraction is given by $a$, the larger is its
value, the larger the oscillations and the slower is the relaxation
to the asymptotic distribution, as is shown in Fig.~\ref{C-t}.
One can notice that the period is roughly the same (same value of
$r_\alpha$), but the amplitude of oscillations is much larger in the
plot to the right (larger $a$).

Since $a=K/N$, these large oscillations make difficult to perform
measurements on the asymptotic state on small populations, but large
values of $N$ require longer simulations. One may say that the model
is intrinsically complex.

The asymptotic cost is given by $C_{\infty}=a\int_0^{v_\alpha}
P_1(\alpha)\propto av_\alpha^{a/r_\alpha}$. As one can
see from Fig.~\ref{alpha}, there is a cost even in the absence of
infection, since the agents have to monitor the level of infection
against the database.
The lower values of the cost are associated to
values of $r_\alpha$
smaller than $a$.

\begin{figure}
 \includegraphics[width=0.6 \columnwidth]{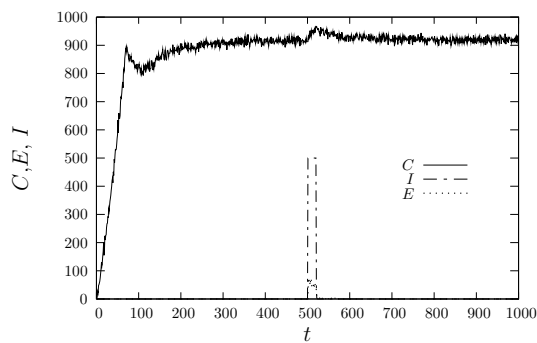}
\caption{\label{infect} Temporal behavior of the cost $C$,
infection $I$ and error level $E$ for $K=5$, $n=500$ ($a=0.01$),
$r=0.005$. The pulse is at the time $500$}
\end{figure}

\section{Infectivity scenarios}\label{scenarios}

The source of infection may be quenched, \emph{i.e.}, a
fraction $p$ of the population always emits tainted messages, or
annealed, in which case the fraction $p$ of the spreaders is
changed at each time step. Let us first study the case of a
pulse of infection (with $p=1$) in the asymptotic state and a
duration $\Delta t = 20$. For large values of the asymptotic cost,
The infection is removed in just a few time steps, as shown in
Fig.~\ref{infect}.

\begin{figure}
 \includegraphics[width=0.5\columnwidth]{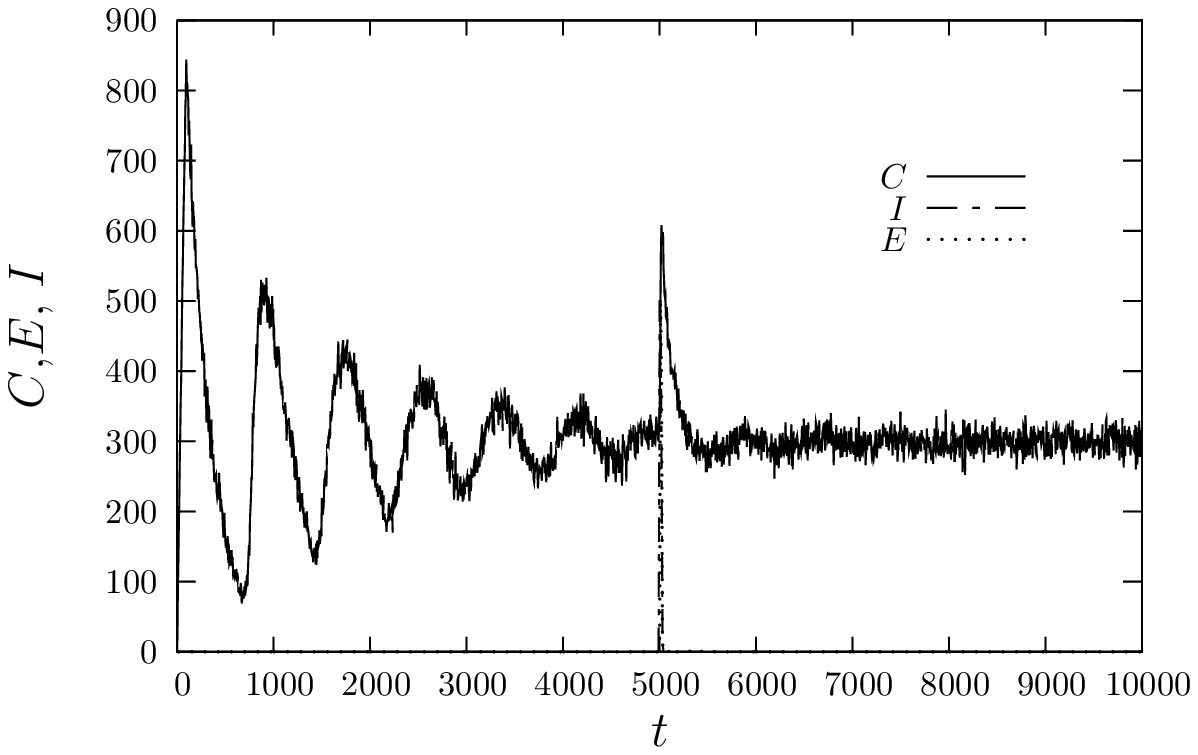}
 \includegraphics[width=0.5\columnwidth]{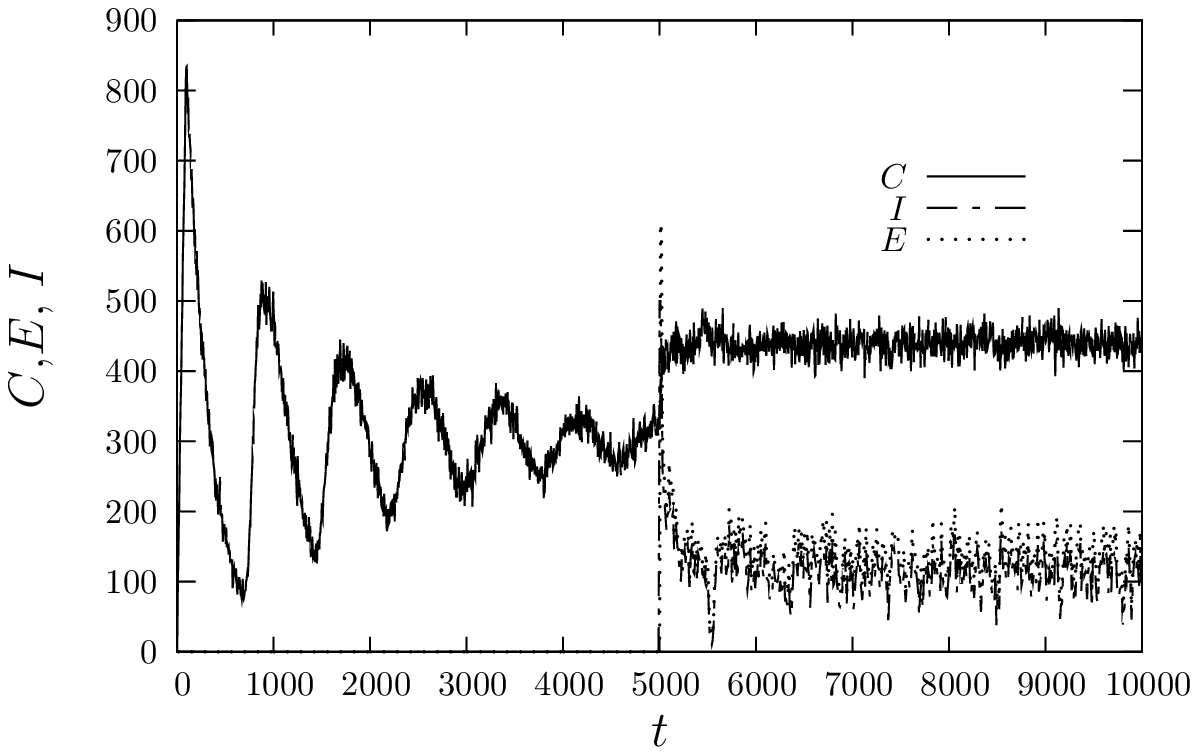}
\caption{\label{Av} Temporal behavior of the cost $C$,
infection $i$ and error level $E$ for $K=2$, $n=500$ ($a=0.004$),
$r=0.001$. Left: $v_A=10^{-3}$ (eradication).
Right: $v_A=10^{-4}$ (endemic infection).  The pulse occurs at time
$5000$}
\end{figure}

For smaller values of the cost, the fate of the infection is related
to the scenario (quenched or annealed infectors). If the infection
level is small, and the infectors are quenched, the rising of the
corresponding $\alpha_{ij}$ efficiently isolate the contagion. In the
case of a ``pulse'' of infection, or for annealed infectors, the fate
of the contagion is mainly ruled by the quantity $A_i$. If $A_i$ grows
rapidly ($v_A$ sufficiently large), a temporary increase of
the cost
is enough to eradicate the epidemics, see Fig.~\ref{Av}. In the
opposite case, the infection becomes endemic even for non-persistent
infectors: it is maintained by the spreading mechanism.
The increment used in the following investigations is small enough
so that we can observe the persistence of the infection.

If the infectors are persistently renewed, the contagion  cannot get
eradicated but only
kept under control. The role of the two heuristics is different in
the two cases.

The representativeness heuristic ($\alpha_{ij}$) is  the optimal
strategy to
detect agents which are constantly less reliable than the others
(quenched case), but it is completely
useless in the annealed case.
The availability heuristic ($A_i$), considering at each time step
the
average infection of the system, is able to control the spread of
infection in the annealed case.

The oblivion mechanism, related to the anchoring heuristic, is a the
key parameter governing the speed of adaptation to variable external
conditions. It controls the oscillations of the cost
(Fig.~\ref{C-t}) and
it is fundamental to minimize the computational load of the control
process. The oblivion of $\alpha_{ij}$ (representativeness
parameter) controls the computational cost at the equilibrium in
both cases. High values
of $r_\alpha$ correspond to a conservative behavior of the
system, in this case a large computational cost and a corresponding
low number of infected and errors characterized the equilibrium. Low
values of $r_\alpha$ correspond to the dissipative behavior for
which the system minimizes the computational cost but allows large
fluctuations of infected and high values of errors.

\section{Dynamical behavior}\label{results}
We run extensive numerical simulations and recorded the asymptotic
cost $C$, number of infected people $I$, and errors $E$ as function
of the oblivion parameters ($r_\alpha$ and $r_A$), the probability
and the pulse of infection ($p$), and the density of contacts ($K$).
In these simulation we kept $v_A=10^{-6}$ in order to stay in the
endemic phase, and therefore $r_A$ did not play any role.

\begin{figure}[htbp]
\vspace*{3.5em}
\centering
\includegraphics[width=6cm]{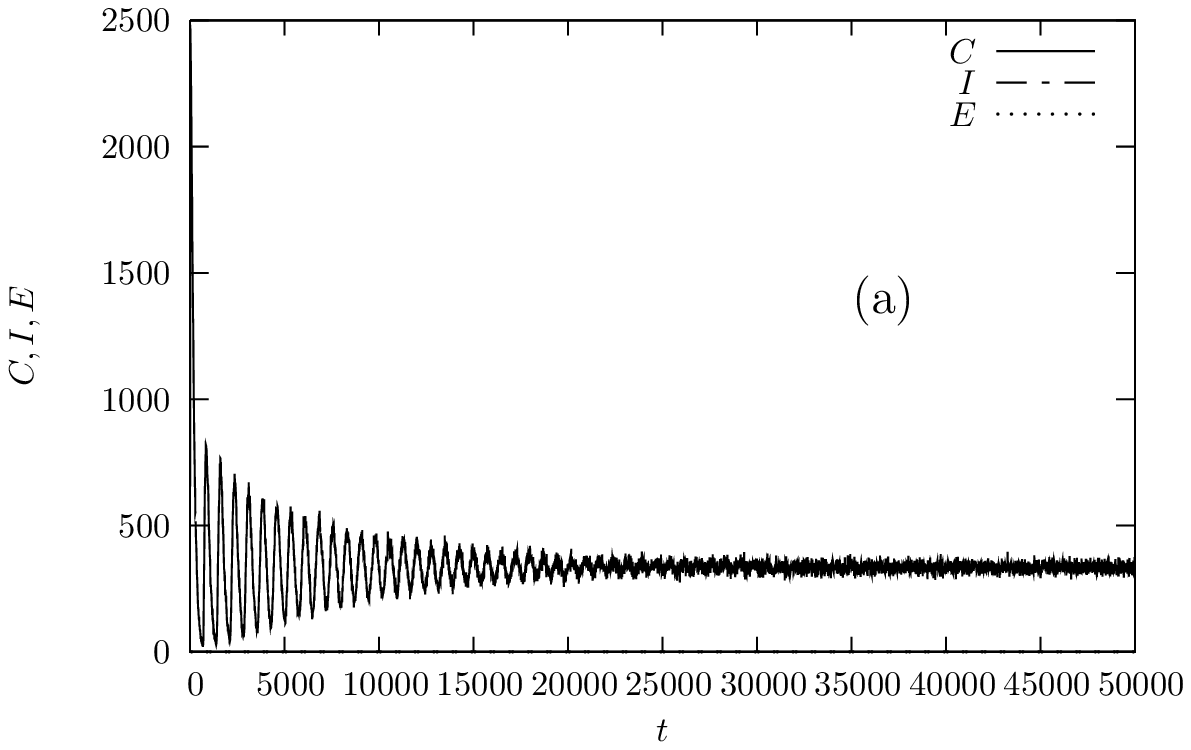}
\includegraphics[width=6cm]{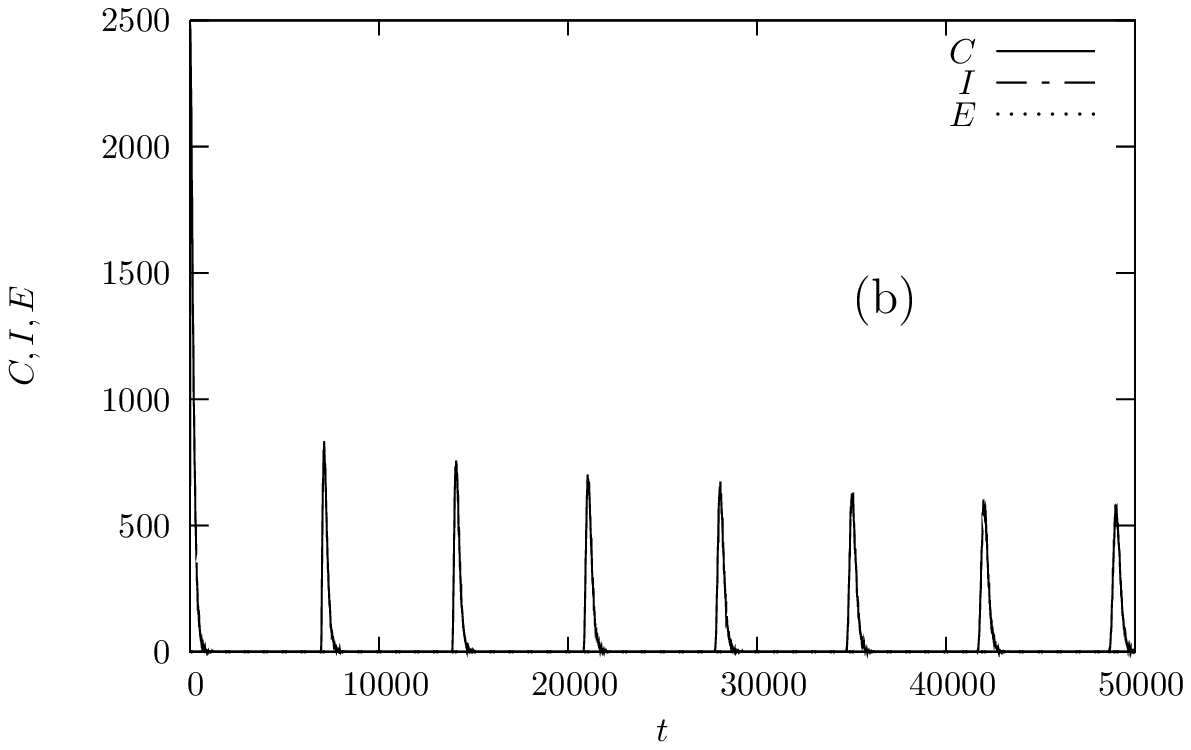}
\includegraphics[width=6cm]{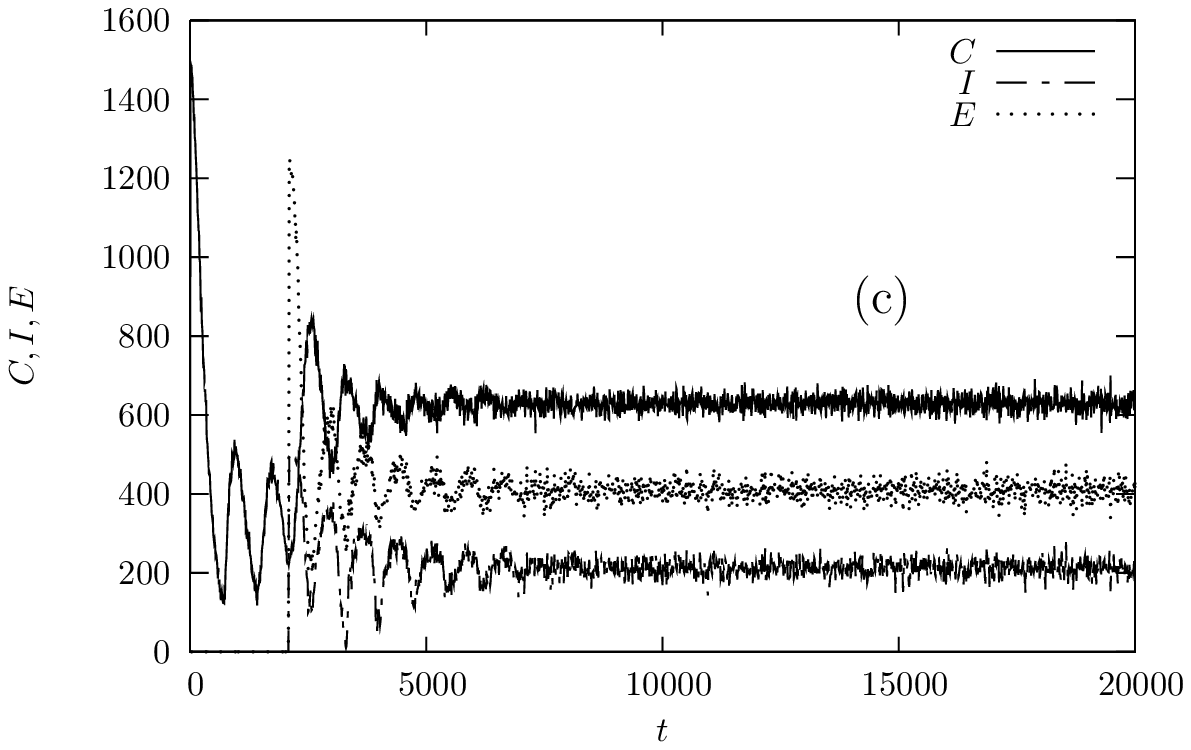}
\includegraphics[width=6cm]{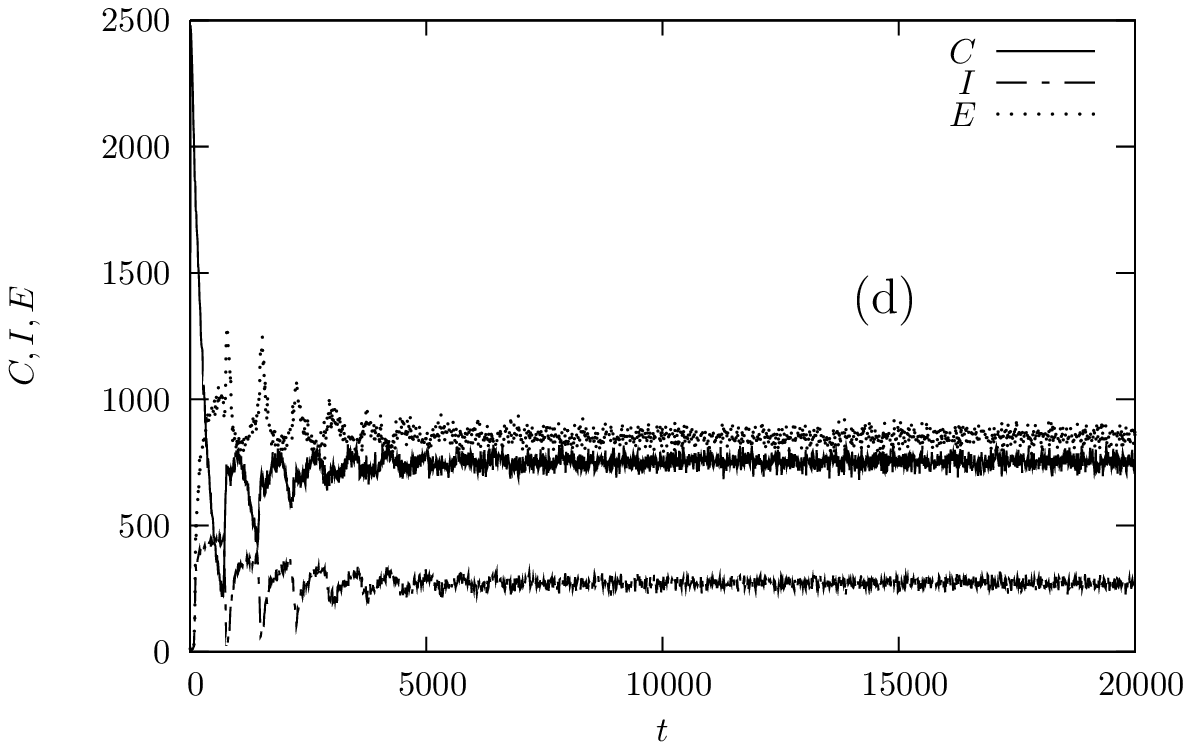}
\includegraphics[width=6cm]{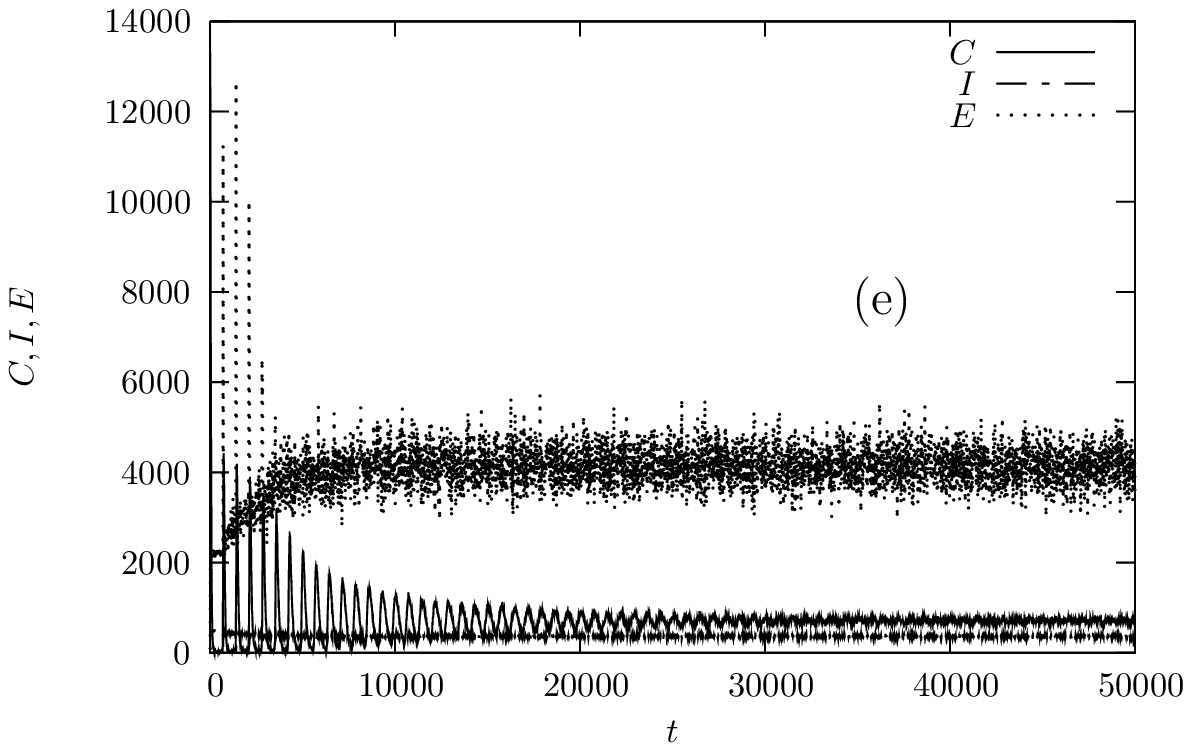}
\includegraphics[width=6cm]{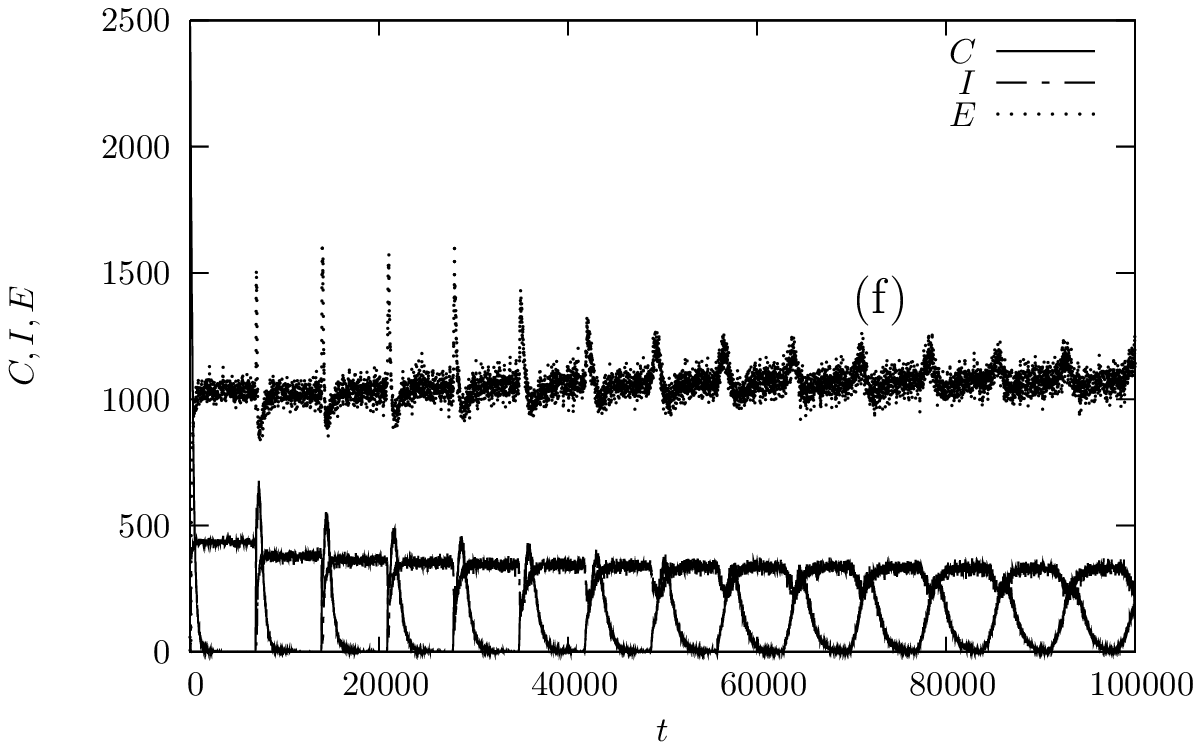}
\caption{\label{infect2} Cost $C$, Infection level $I$ and Errors
$E$ vs. time $t$ for two different values of the parameter
$r_\alpha$, $r_\alpha=10^{-3}$ (a,c,d,e); $r_\alpha=10^{-4}$
(b,f), different values of $p$, $p=0$, (a,b);
$p=10^{-2}$ (d,e,f); $p=10^{-6}$, (c) and for some value of the
connectivity ($K=5$ for plots (a,b,c,d,f), $K=30$ for plot (e)) and
population $N=500$.}
\end{figure}

Fig.~\ref{infect2} shows the effect of infection with different
values of $r_\alpha$ and contact density for for $K=5$  and $N=500$.
Plots (a) and (b) show the oscillatory patterns without infection
($p=0$) for $r_\alpha=10^{-3}$ (a), $r_\alpha=10^{-4}$ (b). The
oblivion $r_\alpha$ (in the presence of infection) changes both the
oscillatory frequency (as studies in the previous section) and the
oscillatory delay before convergence to a basic fluctuation pattern.
Note that increasing $r_\alpha$ the frequency of the oscillations
increases. When $r_\alpha=0.0001$, (a), the period $T$ is
$T=\ln{2} 10^4\approx 7000$; for $r_\alpha=0.001$, (b),
$T\approx 700$). By adding infection (annealed version),  we
obtain a quicker convergence the basal fluctuation equilibrium
(c). We found that the time to reach the basic fluctuation
equilibrium does not depend on the infection probability and the
level of the fluctuation remain unchanged even for long runs (d).
Plots (e) and (f) show that with the same value of the
infection probability, increasing the density of contacts produces
larger fluctuations, a quicker convergence of the cost ($K=30$ for
 plot (e) with respect to $K=5$ for all others). The two scenarios
have different oblivion ($r_\alpha=10^{-3}$ (e), $r_\alpha=10^{-4}$
(f)). Then,
increasing $p$,  the frequency of the oscillations remains the same
but the peaks broaden.

\section{Discussion and Conclusions}\label{conclusions}

In this paper we have been modeled the cognitive mechanisms known as
availability and representativeness heuristics. The role of the first
one in the human decision making process seems to be to produce a
probability estimation of an event based on the relative observed
(registered) frequency distribution.
The second heuristic, representativeness, acts inferring certain
attributes from others easier-to-detect. Both heuristics are liable or
 "noise affected", but surely they represent a very fast way to
analyze environmental data using little quantity of memory and time.
But the very interesting aspect, and not underlined enough in
literature, is the role of the cooperation between heuristics. The
co-occurrence of their activities could be coordinate also in the
human cognition, but of course it is very interesting from a
computational point of view.

We supposed that the  availability heuristic corresponds to a mean
field estimation of the ``risk'', while representativeness partially
maintains the memory of the previous interactions with the others.
In the quenched and annealed scenarios we can capture the effect of
the heuristics coordination. The quenched scenario considers the case
of ``systematic spreaders'' where same agents emits at each time step
a tainted message. In this case the availability heuristic would fails
to minimize cost and infection if representativeness was absent.

On the contrary in the annealed scenario the spreaders are completely
chosen at random at each time step. In this extreme case there is no
information contained in the previous history of the system, and
representativeness heuristic became completely useless. In this
situation the only available information is the rate of infection,
and availability heuristic is the most efficient way to minimize both
cost and risk of infection.

The oblivion mechanism associated to the two heuristics determines
both the cost of the control process and a sort of its reactivity. In
average the cost, which represents the number of
operations/computation to cope the task, is proportional to the
oblivion value, the number of infected and errors are inversely
proportional to the oblivion. If the cost as so as it happens in the
biological domain, is considered as a quantity which the system has to
minimized, it means that will exist an optimal value of both oblivion
parameters for each possible condition.
The reactivity of the control process could be defined as the time
needed from the system to reduce to zero a new infection. In our model
the oblivion of both the two heuristics appears to control also the
size of ``cost oscillations''. We found that the larger the oblivion
level, the lower the oscillations and the time needed to reach the
asymptotic equilibrium.

Our simulations show that under the infection, the cost reaches its
asymptotic value much  earlier than without infection. This suggests
that a low value of infection level may even provide  some
advantages for the quick dumping of the oscillatory behavior
resulting in an improved cost predictability.

The investigation of heuristics exploits a major overlap between
artificial intelligence (AI), cognitive science and psychology. The
interest in heuristics is based on the assumption that humans process
information in ways that computers can emulate and heuristics may
provide the basic bricks for bridging from brains to computers . Our
model framework approach is quite general and offers some points of
reflections on how the study of complex systems may become help
developing new areas of AI.
In the past years the AI community has debated as to whether the mind
is best viewed as a network of neurons (connectionism), or as a
collection of higher-level structures such as symbols, schemata,
heuristics, and rules, \emph{i.e.}, emphasizing the role of symbolic
computation. Nowadays the symbolic representations to produce general
intelligence is in slightly decline but the ``neuron ensemble''
paradigm has also shifted towards more complex models particularly
taking into account and combining findings from both fNMR and
cognitive psychology fields (~\cite{Gazzaniga2002,Rosenzweig2004}).

Here we show that the incorporation of simple heuristics in a small
network of agents leads to a rich and complex dynamics.Our model does
not take into account mutation and natural selection which is of key
importance for the emergence of complex behavior in animal societies
and in the brain development (see for example Pinker and the follow up
debate ~\cite{Pinker1997}).

A multi-agent model, where each agent represent a message/modifying
person/neuron, can serve as a very natural abstraction of
communication networks, and hence be easily used by psychologists as
well as computer scientists. Such a model also allows the tracking
of single agent fates so that communities with low member numbers
are easily dealt with and these models also provide for much more
detailed analysis compared to average population approaches like
continuous differential equations.

Heuristics may have even greater value in case of environmental
challenges, i.e. organisms need to adapt quickly to environmental
fluctuations, for example starvation and high competition, they must
be able to make inferences that are fast, frugal, and accurate.
These real-world requirements lead to a new conception of what
proper reasoning is: ecological rationality. Fast and frugal
heuristics that are matched to particular environmental structures
allow organisms to be ecologically rational. The study of ecological
rationality thus involves analyzing the structure of environments,
the structure of heuristics, and the match between them.

\end{document}